\documentstyle[12pt,epsfig]{article}                                
\begin{document}
\titlepage                  
\vspace{0.5in}
\begin{center}
\begin{Large}
\begin{bf}

 Revisiting "Morse Potential on Quantum Computer for Molecules and Supersymmetric Quantum Mechanics" :arXiv:2102.05102v1[quant-ph]

\vspace{0.1in}

 Biswanath Rath 

\vspace{0.1in}
\end{bf}

\end{Large}

Department of Physics, Maharaja Sriram  Chandra  Bhanja Deo University, Baripada,Orissa,India.

\vspace{0.1in}

E.mail:biswanathrath10@gmail.com

\end{center}
\begin{bf}
Abstract: 
\end{bf}

We find the theoretical results  on energy eigenvalues and corresponding 
supersymmetric Hamiltonians reflect contradictory behaviour for negative
 values of A. Furthermore  the resulting susy partner potentials can be a model for scattering 
states instead of bound states. However following the literature( Flugge(1979);  Jafarpour and Afshar(J.Phys A (2020)), we suggest a correct form of superpotential, which remains valid for both  positive or negative values of constants. Apart from this in complex space also the eigenvalues remain invariant without the discussion of T-symmetry,  as the previous discussion appears incomplete.

\begin{bf}
1.Introduction
\end{bf}

 We believe Morse potential[1] is as old as quantum mechanics. Nearly nine decades ago Prof P.M.Morse suggested the potential to study spectral nature of molecules. Later on, in 1982 Witten[2] used supersymmetry to study this model potential. Even standard problems in quantum mechanics also address it elaborately[3].
In (2002) Jafarpour et.al [4] revisited the model in calculating energy eigenvalues using a modified vacuum operator analysis. In fact without proper review of the literature, Apanavicius,Feng,Flores, Hassan and McGuigan in an arxiv 
communication[5] have suggested supersymmetric quantum mechanics using Morse 
 oscillator and suggested an
 expression for energy level 
\begin{equation}
E_{n}=E_{n}^{(-)}= A^{2} - (A-n)^{2}
\end{equation}
for the superpotential 
\begin{equation}
W(x)= A - e^{-x}
\end{equation}
It is seen that the energy level expression for $A\rightarrow - A$ becomes 
\begin{equation}
\in_{n}= A^{2} - (A+n)^{2}
\end{equation}
Hence for the ground state still has zero energy i.e
\begin{equation}
\in_{-}= 0
\end{equation}
Let us write the SUSY partener Hamiltonians as 
\begin{equation}
H^{-}= p^{2} +  e^{-2x} - (1+2A) e^{-1} + A^{2}       \rightarrow (A= + ve)
\end{equation}
becomes 
\begin{equation}
h^{-}= p^{2} +  e^{-2x} - (1-2A) e^{-x} + A^{2}       \rightarrow (A= - ve)
\end{equation}

For any value of $A = -1,-2,-3,-4,-5,-6,-7,-8,-9,-10,....$, the  above 
operator ( $h^{-)}$ )will no longer hold any bound states. However it becomes 
a good model[5] for scattering study[6].In other words $h^{(-)}$ is a 
scattering Hamiltonian.
Now let us consider, the other partner hamiltonian 
\begin{equation}
H^{+}= p^{2} +  e^{-2x} - (2A-1) e^{-1} + A^{2}       \rightarrow (A= + ve)
\end{equation}
For negative A the partner Hamiltonian becomes 
\begin{equation}
h^{+}= p^{2} +  e^{-2x} + (1+2A) e^{-x} + A^{2}       \rightarrow (A= - ve)
\end{equation}
This is also fit for scattering study. So the model reported earlier[5] is no longer suitable for a generalisation of supersymmetric study.
 Before we rectify the above anomaly, we proceed as follows.

\begin{bf}
2. Morse model potential[2,3] 
\end{bf}

Let us consider the Morse model Hamiltonian as[2,3] 

\begin{equation}
H_{Morse}=\frac{p^{2}}{2}+D[1-exp(-\lambda x)]^{2}
\end{equation}
having energy level as 
\begin{equation}
E_{n}=\lambda \sqrt{2D} [(n+\frac{1}{2})-(n+\frac{1}{2})^{2} \frac{\lambda}{\sqrt{8D}}]
\end{equation}

Interestingly using matrix diagonalisation method [7-10], we verify the same.
 In other words numerical values of MDM clearly match  with analytical results.

\begin{bf}
3.Supersymmetric models using Morse potential 
\end{bf}

Let us briefly describe key points in supersymmetry[11,12]

\begin{equation}
 H_{\mp}= p^{2} + W^{2} \pm \frac{dW}{dx}
\end{equation}
with where super potential $W(x)$ satisfies the relation 
\begin{equation}
 B_{-}\Psi_{0}= \frac{d \Psi_{0}}{dx} + W(x)\Psi_{0}=0
\end{equation}
The corresponding energy levels must satisfy the relations 
\begin{equation}
E_{0}^{(-)}=0
\end{equation}
and 
\begin{equation}
E_{n+1}^{(-)} = E_{n}^{(+)}
\end{equation}

Now we consider the superpotential ,$W(x)$ as 
\begin{equation}
 W(x)= V (1-e^{-x})
\end{equation}
In this case the ground state wave function is determined by the above 
e condition is found to be 
\begin{equation}
\Psi_{0}(x)\sim e^{-x - e^{-x}}
\end{equation}
This wave function is well behaved i.e 
\begin{equation}
\Psi_{0}(x\rightarrow \infty)\rightarrow 0
\end{equation}
\begin{equation}
\Psi_{0}(x\rightarrow -\infty) \rightarrow 0
\end{equation}
Now using MDM we find the energy levels for $V=\pm 10$ as 
\begin{equation}
E_{n}^{(-)}=0;19;36;64;..
\end{equation}
and 
\begin{equation}
E_{n}^{(+)}=19;36;64;..
\end{equation}
Interested readers will find that SUSY energy conditions are satisfied and 
 independent of sign of V. However one has to interpret it accordingly by using the relation 
\begin{equation}
E_{n+1}^{(-)}=E_{n}^{(+) }(A\rightarrow +ve) ; E^{(+)}_{n+1}=E_{n}^{(-)} (\rightarrow A=-ve)
\end{equation}

\begin{bf}
4.Supersymmetric models using Morse potential in complex space  
\end{bf}

Let use use similarity transformation as 
\begin{equation}
 S x S^{-1}= x + ip 
\end{equation}

\begin{equation}
 S p S^{-1}= p
\end{equation}

with 
\begin{equation}
 W(x)=V(1- e^{-x-ip})
\end{equation}
 Under this transformation, SUSY operators are non-Hermitian and satisfies the relation
\begin{equation}
 [H^{(\pm)},T]=0
\end{equation}
where T-stands for time reversal operator having the properties :$T x T^{-1}$;$Ti T^{-1}=-i$ and $ T p T^{-1}= -p $.For details [6,7]. 

\begin{bf}
5. Conclusion
\end{bf}

In this visit to  SUSY, our suggested model becomes a better choice as compared to earlier[5]. We have just pointed to a few anomalies in previous work[5].Hope authors[5] and interested readers will note of this presentation before any 
 possible publication involving SUSY using Morse potential. We also suggest a complex form of supersymmetric Morse potential, whose energy levels remain invariant under T-symmetry condition also. Hope authors will focus attention on the new symmetry without which previous work remains incomplete.

\end{document}